\documentstyle[preprint,tighten,aps]{revtex}
\begin{document}
\draft
\preprint{IFUP-TH 42/96}
\title{
Classical improvement of lattice actions and quantum effects: \\
a unified view.
}
\author{Paolo Rossi and Ettore Vicari}
\address{
Dipartimento di Fisica dell'Universit\`a and I.N.F.N.,
I-56126 Pisa, Italy
}
\maketitle

\begin{abstract}
The possibility of removing the one-loop perturbative effects of
lattice  artifacts
by a proper choice of the lattice action is explored, and found to 
depend crucially on the properties of the physical quantity considered.
In this respect
the finite-space-volume mass gap $m(L)$ is an improved observable.
We find an explicit momentum space representation of the one-loop 
contribution
to $m(L)$ for arbitrary lattice actions in the case of two-dimensional
$O(N)$ $\sigma$ models. We define a ``tree perfect'' Symanzik action 
and find
that it formally removes all one-loop lattice artifacts in $m(L)$.
On-shell improved actions do not share this property.
\end{abstract}



Improvement of the lattice action has been often advocated 
as a possible way out of the problem of finite lattice spacing effects,
that obscure the scaling properties in numerical simulations on
small lattices. The original idea can be traced back to
Wilson~\cite{Wilson}, 
who introduced
the notion of renormalization group trajectory. Lattice actions 
determined 
by the trajectory of a renormalization group transformation are free
of lattice artifacts. Due to the difficulty of effectively
finding a renormalized trajectory,
two major strategies of improvement were suggested
by Symanzik~\cite{Symanzik} and by Hasenfratz and Niedermayer~\cite{HaNi}.
They differ both in purpose and in many technical details, 
but they share the possibility of defining a ``classically perfect
action'' whose properties may then be compared.
By ``classically perfect action'' we mean a lattice action 
which does not present any power-like dependence on the lattice spacing
in the tree evaluation of spectral properties, and of properly
chosen correlation functions.

In Symanzik's case, this action is of no practical relevance, since
it involves infinitely-long-range interactions.
However its properties are theoretically
interesting in view of a deeper understanding of the whole improvement
program. 

By renormalization group arguments, one can infer that
quantum effects are more important
than higher order classical finite lattice spacing contributions. It is
therefore worth exploring the mechanism that might lead to important
cancellations of these quantum effects. Symanzik's
``tree perfect'' action constitutes a reasonably simple laboratory
in order to check some ideas that have been put forward.

Hasenfratz and collaborators~\cite{FaHaNiPa} stressed that
in two-dimensional $O(N)$ $\sigma$ models
their version of a ``classically perfect'' action showed
no power-like dependence on the lattice artifacts in  the one-loop
contribution to the finite-space-volume mass gap.
Such lattice artifacts are instead
found when calculating the same quantity with the standard nearest-neighbor
action. The same authors already observed that the above effects were
consistently reduced in the simplest (second-nearest-neighbor) version
of  Symanzik improvement. 
Their approach was however intrinsically numerical,
and therefore not easily adapted to exploring the different possibilities 
in full generality. We therefore
decided to move a step forward, and to look for an analytic
expression of the one-loop contribution to the finite-space-volume mass gap 
in two-dimensional $O(N)$ $\sigma$ models, that could hold
for an arbitrary form of the lattice action, allowing a systematic study 
of finite-lattice effects.

We were greatly helped in our effort by the friendly collaboration
of M.~L\"{u}scher and P.~Weisz, who generously made their private
notes available to us. 

For definiteness, let us recall that any reasonable definition of 
finite-space-volume mass gap $m(L)$
($L$ is the size of the space volume)
in 2-d $O(N)$ $\sigma$ models admits a loop expansion in the
form~\cite{Luscher}\cite{LuWeWo}:
\begin{equation}
{m(L)L} = {N-1\over 2} g^2 \sum_{l=0}^\infty A_l(L) g^{2l}
\label{eq1}
\end{equation}
where $l$ is the number of loops and $A_l(L)$ in turn have an
asymptotic  expansion in powers of $a^2/L^2$:
\begin{equation}
A_l(L) = \sum_{n=0}^\infty 
\left( {a^2\over L^2}\right)^{n} \sum_{p=0}^l a^{(n)}_{lp} 
\left( \ln {L\over a}\right)^{l-p}.
\label{eq2}
\end{equation}
We are neglecting possible exponentially depressed finite-lattice
spacing corrections.
The coefficients $a^{(n)}_{lp}$ are in general dependent on $N$.
Recursive relations between the coefficients $a_{lp}^{(n)}$
($n$ fixed and different from zero) are dictated by the renormalization 
group properties of the higher dimensional operators, whose effective 
presence in the lattice action generates 
the scaling violations to $O\left[ (a^2/L^2)^{n}\right]$~\cite{CuMePa}.
Terms with $n=0$, i.e. 
the limit $a^2/L^2\rightarrow 0$ of  $m(L)L$, define a 
running coupling constant $\bar{g}$~\cite{LuWeWo},
\begin{equation}
\bar{g}^2=
g^2\sum_{l=0}^\infty \sum_{p=0}^l g^{2l} a^{(0)}_{lp} 
\left( \ln {L\over a} \right)^{l-p},
\label{eq3}
\end{equation}
obeying a homogeneous renormalization group equation.

It is convenient to reexpress the general result into the form
\begin{equation}
m(L)L= {N-1\over 2}g^2 \sum_{n=0}^\infty
\left( {a^2\over L^2}\right)^{n} \sum_{p=0}^\infty 
g^{2p} b_p^{(n)}\left( g^2\ln L/a\right)
\label{eq4}
\end{equation}
where 
\begin{equation}
b^{(n)}_p(x)=\sum_{q=0}^\infty a^{(n)}_{p+q,p}x^q
\label{eq5}
\end{equation}
are completely determined by $a_{pp}^{(n)}$ and the RG recursion
equations.

Symanzik's improvement program may act separately on the indices 
$p$ and $n$, which are respectively related to higher-loop 
contributions and to insertions of
higher-dimensional operators. 
In the usual language, tree-improvement would
consist in removing the coefficients $a_{00}^{(n)}$ for $n\ne 0$, 
which automatically
would also imply removing all coefficients $a^{(n)}_{q0}$,
i.e. setting $b^{(n)}_0=0$.
In turn, one-loop improvement would amount to setting
$a^{(n)}_{11}=0$, that is $b_1^{(n)}=0$.
However, this point of view would be misleading for the problem at hand. 
The basic reason for this statement stays in  fact
that the condition $a_{00}^{(n)}=0$ for $n\ne 0$
(i.e. $A_{0}(L)= 1$) is obtained automatically, for any action, 
by a proper choice of the definition of $m(L)$, making no reference 
whatsoever to any  improvement program.

The finite-space-volume mass gap
$m(L)$ can be defined as the coefficient of the exponential decay of the
wall-wall correlation function in the time direction, i.e. 
for $|x_0-y_0|\rightarrow \infty$
\begin{equation}
{1\over L^2}\sum_{x_1,y_1}\langle s_x \cdot s_y\rangle
\sim \exp \left[ - m(L) |x_0-y_0|\right].
\label{eq6}
\end{equation}
Exponential decay in perturbation theory is insured 
by taking free boundary conditions in 
time, and periodic boundary conditions in space
($L$ points in a circle)~\cite{Luscher}~\cite{LuWeWo}.

As we shall see, $m(L)$ enjoys the properties that
$A_0(L)=1$ in the expansion (\ref{eq1}) independently
of the choice of lattice action.
The renormalization group then predicts the one-loop result
$a_{10}^{(n)}=0$ for $n\ne 0$, i.e. absence of terms proportional 
to $\ln ({L/ a})$
in the finite lattice one-loop corrections to the mass gap.

Which whould be in this case the effect of a tree-improvement, if any?
As one might have conjectured Symanzik tree-improvement will now act 
on the coefficients $a_{11}^{(n)}$, setting it to zero. As
a consequence we would have $b_1^{(n)}=0$ as a tree-improvement
effect, without
the need to appeal to any notion of one-loop quantum
improvement. 

In order to produce evidence for these statements, let us
consider the explicit expression of the one-loop contributions
to $m(L)L$. At the one-loop level, full generality is achieved
starting  from a lattice 
action whose parametrization is the following~\cite{HaNi}:
\begin{equation}
{\cal A} = -{1\over 2}\sum_{n_1n_2} \rho_{n_1n_2}
\left(1 - s_{n_1}\cdot s_{n_2}\right)
+ \sum_{n_1n_2n_3n_4} c_{n_1n_2n_3n_4}
\left(1 - s_{n_1}\cdot s_{n_2}\right)
\left(1 - s_{n_3}\cdot s_{n_4}\right).
\label{eq8}
\end{equation}
$n_i$ are lattice sites, and the functions $\rho_{n_1n_2}$ 
and $c_{n_1n_2n_3n_4}$ enjoy the following exchange symmetries
\begin{equation}
\rho_{n_1n_2}=\rho_{n_2n_1},
\label{eq9}
\end{equation}
\begin{equation}
c_{n_1n_2n_3n_4}=c_{n_2n_1n_3n_4}=c_{n_1n_2n_4n_3}=c_{n_3n_4n_1n_2}.
\label{eq10}
\end{equation}
Possible higher order interactions in the lattice
action do not contribute at one-loop.
We shall not assume translation invariance, at least
in the time direction, since it is in general violated by
arbitrary boundary conditions.

We must now compute the two-point spin-spin correlation function
up to $O(g^4)$ without relying on any specific feature of the action
and for arbitrary boundary conditions.
We follow the coordinate space approach~\cite{Luscher} in the version
suggested by Cline~\cite{Cline}.
We introduce the massless propagator $P_{xy}$ which is a solution of the
difference equations
\begin{equation}
\sum_z \rho_{xz} P_{zy}=\delta_{xy} - {1\over V},
\label{eq11}
\end{equation}
where $V$ is the total lattice volume. We have removed the zero-mode
by appropriate gauge-fixing~\cite{Hasenfratz}.
$P_{xy}$ will not in general be translation-invariant,
as a consequence of the non-invariance of $\rho_{xy}$.

The resulting coordinate-space two-point function is
\begin{eqnarray}
\langle s_x\cdot s_y\rangle = &&
1 + g^2 (N-1) G_{xy} \nonumber \\
&& + g^4(N-1){1\over 2} \left[ G_{xy}^2+\sum_{n_1n_2}
\left( \rho_{n_1n_2}G_{n_1n_2}-\delta_{n_1n_2} \right)
\Delta G_{xy,n_1}\Delta G_{xy,n_2} \right]\nonumber \\
&&+ g^4 (N-1)^2{1\over 4}\sum_{n_1}
\left( {2\over V} + \sum_{n_2} \rho_{n_1n_2}  P_{n_2n_2}\right)
\left(\Delta G_{xy,n_1}\right)^2 \nonumber \\
&& + g^4 (N-1) 4 \sum_{n_1n_2n_3n_4} c_{n_1n_2n_3n_4} 
\left( \Delta G_{xy,n_1} - \Delta G_{xy,n_2} \right)
\left( \Delta G_{xy,n_3} - \Delta G_{xy,n_4} \right) G_{n_1n_3}
\nonumber \\
&&-g^4(N-1)^2 \sum_{n_1n_2n_3n_4} c_{n_1n_2n_3n_4} 
\left( \Delta G_{xy,n_1} - \Delta G_{xy,n_2} \right)^2 G_{n_3n_4},
\label{eq12}
\end{eqnarray}
where we have introduced the notations
\begin{eqnarray}
&&G_{xy} = P_{xy} - {1\over 2}\left( P_{xx}+P_{yy}\right),\nonumber \\
&&\Delta G_{xy,n}= G_{xn}-G_{yn} .
\label{eq13}
\end{eqnarray}
It is worth noticing that in the case of open boundary conditions
$G_{xy}$ is time-translation invariant up to terms
that are exponentially depressed in the time distance $|x_0-y_0|$.

When considering space translation invariance, say by taking periodic
boundary conditions in space, one can actually perform a Fourier 
transform of the space coordinate and reduce the problem to an effective
one-dimensional equation in time. In the case of open boundary
conditions, only the zero-space-momentum component of the propagator
will be seriously affected by the lack of translation invariance;
all other components will have violations of invariance
that are exponentially depressed with the time distance~\cite{LuWeWo}.

In view of the above considerations, 
one may avoid solving the 
one-dimensional problem with free boundary
conditions for an arbitrary form of the action. Indeed the solution for
the most general action (\ref{eq8}) turns out to differ
from that of the standard (nearest-neighbor) action
only by terms that are exponentially depressed in time, and therefore
not sensitive to the boundaries.
As a consequence we may express the general result for $m(L)L$
in terms of the standard action result plus terms that can be computed
with periodic 
boundary conditions, hence directly in momentum space.

Without belaboring on the details of the derivations, we may express
our final result for $m(L)$ in terms of the following Fourier
transforms:
\begin{equation}
\widetilde{\rho} (p)={1\over V} \sum_{n_1n_2}\rho_{n_1n_2}
e^{-ip\cdot (n_1-n_2)},
\label{eq14}
\end{equation}
\begin{equation}
\widetilde{c}(p,q,r)={1\over V^3} \sum_{n_1n_2n_3n_4}c_{n_1n_2n_3n_4}
e^{-ip\cdot (n_1-n_2)}e^{-iq\cdot (n_3-n_4)}
e^{-ir\cdot\case{1}{2}(n_1+n_2-n_3-n_4)},
\label{eq15}
\end{equation}
where complete symmetrization of the dummy indices, space periodicity 
and the limit $T\rightarrow \infty$ are assumed.
Notice that $\widetilde{c}(p,q,r)$ enjoys also the following symmetry
$\widetilde{c}(p,q,r)=\widetilde{c}(q,p,r)$.
We introduce the parametrization
\begin{equation}
A_1(L) = r_1(L) + (N-2)r_2(L) + s_1(L) + (N-1)s_2(L),
\label{eq17}
\end{equation}
where $s_i$ are vanishing when $c_{n_1n_2n_3n_4}=0$.
We obtained
\begin{equation}
r_1(L)= -{1\over L} \sum_{l=1}^L
\int {dk_0\over 2\pi} {1\over \widetilde{\rho}(k_0,k_1)}
\left[ {1\over 2} {\partial^2\over \partial k_0^2}
\widetilde{\rho}(k_0,k_1) - 1\right],
\label{eq18}
\end{equation}
\begin{equation}
r_2(L)= {1\over L} \left[ \sum_{l=1}^{L}
\int {dk_0\over 2\pi} {1\over \widetilde{\rho}(k_0,k_1)}
-\int {dk_0\over 2\pi} {1\over \hat{k}_0^2} \right],
\label{eq19}
\end{equation}
where the different dependence on the
(continuous) time component $k_0$ and the (discrete)
space components 
\begin{equation}
k_1={2\pi l\over L},\;\;\;\;\;\;\;\; l=1,...L
\label{eq20}
\end{equation}
is made explicit, and
\begin{equation}
\hat{k}^2=4\sin^2 (k/2)
\label{eq20b}
\end{equation}
 is the standard action
momentum space representation of the inverse propagator.
Moreover we obtained
\begin{equation}
s_1(L) = -{1\over L} \sum_{l=1}^L
\int {dk_0\over 2\pi} {8\over \widetilde{\rho}(k_0,k_1)}
\left[ {\partial \over \partial q_0}{\partial \over \partial q'_0}
c(q_0,0;q'_0,0;k_0,k_1)\right]_{q_0=q'_0=k_0/2},
\label{eq21}
\end{equation}
\begin{equation}
s_2(L) = {2\over L} \sum_{l=1}^L
\int {dk_0\over 2\pi} {1\over \widetilde{\rho}(k_0,k_1)}
\left[ {\partial^2 \over \partial q_0^2}c(q_0,0;0;0)-
{\partial^2 \over \partial q_0^2}c(q_0,0;k_0,k_1;0)\right]_{q_0=0}.
\label{eq22}
\end{equation}

Since the low-momentum
behavior of $\widetilde{\rho}$ is dictated by the continuum limit
to be
\begin{equation}
\widetilde{\rho}(k_0,k_1)\longrightarrow k_0^2+k_1^2 + O(k^4)
\label{eq23}
\end{equation}
and the low-momentum behavior of $\widetilde{c}$ 
is assumed to be a regular function of $k$,
in the large $L$ limit
the only singular dependence on $L$ is expected to be
originated  from $r_2(L)$.
This singularity is simply parametrized by
\begin{equation}
r_2(L)={1\over 2\pi}\ln {L\over a} + {\rm regular\;\;terms},
\label{eq24}
\end{equation}
as predicted by our renormalization group considerations.

We tested our expressions of $r_i(L)$, $s_i(L)$ against explicit 
(finite $L$, finite $T\gg L$) evaluation of the one-loop contribution
to $m(L)L$ (i.e. using Eq.~(\ref{eq12})).
The propagator $P_{xy}$ wad found by numerically solving Eq.~(\ref{eq11}).
We found full agreement for various forms of the action, including
Symanzik off-shell~\cite{MaPaPe} and on-shell~\cite{LuWe} improved
versions and mixed $O(N)$-$RP^{N-1}$ models. The correct large-$N$ limit is 
also reproduced.
Furthermore the $L\rightarrow \infty$ limits of $r_i$, $s_i$ may be 
shown to agree in general with the computation of Ref.~\cite{HaNi} of 
the $\Lambda$-parameter ratios. 

Our expressions allow a rather direct
testing of the effects that an arbitrary choice of the action may have
on the finite-space-volume, finite-size-scaling-violation effects 
at the one-loop level. They are especially useful in order to test the
effects of a systematic tree-level 
Symanzik improvement. 
To this purpose we note that within Symanzik's program
the quadrilinear couplings $c_{n_1n_2n_3n_4}$ are absent, and
a systematic tree-level improvement can be achieved by choosing 
\begin{equation}
\widetilde{\rho}(k_0,k_1)=\sigma_n(k_0) + \sigma_n(k_1),
\label{eq25}
\end{equation}
where the index $n$ is related to the tree-level Symanzik's improvement
degree. The family of inverse one-dimensional propagators
$\sigma_n(k)$ is defined by
\begin{equation}
\sigma_n(k)=\sum_{t=1}^n {2\over t^2} {(t!)^2\over (2t)!}
\left( \hat{k}^2\right)^t, 
\label{eq26}
\end{equation}
enjoying the basic property
\begin{equation}
{1\over 2} {\partial^2 \sigma_n(k)\over \partial k^2}=1 - 
{(n!)^2\over (2n)!}
\left( \hat{k}^2\right)^n 
\label{eq27}
\end{equation}
For $n=1$ we recover the standard action.

By substituting the above relationships into the expression of 
of $r_i(L)$ (cfr. Eqs.~(\ref{eq18}) and (\ref{eq19}))
one finds that, for $L$ sufficiently larger than $n$, 
\begin{equation}
r_1^{(n)}(L)  = \bar{r}_1^{(n)}+O\left[ \left(a^2/L^2\right)^{n}\right],
\label{eq28}
\end{equation}
\begin{equation}
r_2^{(n)}(L)  = {1\over 2\pi}\ln {L\over a}+\bar{r}_2^{(n)}+
O\left[ \left(a^2/L^2\right)^{n}\right],
\label{eq29}
\end{equation}
where $\bar{r_i}^{(n)}$ are constants that can be
computed in the infinite space-volume limit. 
Corrections are power-series 
expandable in the powers of $a^2/L^2$.
The numerically evaluated coefficients of the leading-order power correction
turn out to  rapidly grow with increasing $n$, suggesting some non-uniformity
in the large-$n$ limit.

The above results exhibit explicitly the correct continuum RG one-loop
coefficient, 
the cancellation of all dependence on $\ln (L/a)$ in finite lattice 
effects and 
the cancellation of all $(a^2/L^2)^{t}$ dependence for $t<n$ in the $n$-th 
order tree-improved action.
This is exactly the pattern described in our general analysis.
The action defined by the (formal) limit $n\rightarrow \infty$ is what
we call the ``Symanzik tree perfect'' action. It essentially amounts to a 
``SLAC derivative'' lattice version of the continuum action,
and shows no power-like corrections to scaling 
when the limit $L\rightarrow\infty$ is taken before the limit 
$n\rightarrow\infty$.
One must not however forget that, since this action involves
infinite-range correlations, at finite $L$ there are finite-volume
effects related to 
the unavoidable truncation. These effects appear as $O(1/L^2)$ corrections
when $n\gg L$.
We insist that the relevance of the result 
is purely conceptual, in that it shows that tree-improvement may lead
in this case to cancellation of one-loop finite-volume effects.

The ``Symanzik tree-perfect'' action allows exact evaluation of some
$n\rightarrow\infty$ limits, that are obtained by setting
$\sigma_\infty(k)=k^2$ and 
integrating between $-\pi$ and $\pi$ in the $k_0,k_1$ variables.
We obtained the following analytical results:
\begin{eqnarray}
\bar{r}_1^{(\infty)} &=& {1\over 4\pi},\nonumber \\
\bar{r}_2^{(\infty)} &=& {1\over 2\pi}
\left( \gamma_E-\ln 2-{2\over \pi} G\right),
\label{eq30}
\end{eqnarray}
where $G$ is Catalan's constant.
The $n\rightarrow\infty$ limit is reached with $O(n^{-1/2})$ corrections in
the case of $\bar{r}_1$, and $O(n^{-1})$ corrections in the case of 
$\bar{r}_2$. For comparison we recall that
for the stardard nearest-neighbour action~\cite{Luscher}:
\begin{eqnarray}
\bar{r}_1^{(1)} &=& {1\over 4},\nonumber \\
\bar{r}_2^{(1)} &=& {1\over 2\pi}
\left( \gamma_E+{1\over 2}\ln 2-\ln \pi\right).
\label{eq31}
\end{eqnarray}

Because of the recent upsurge of interest in the so-called 
``on-shell improved''
lattice actions, we also considered the simplest representative
of this class of actions in the context of two-dimensional $O(N)$
$\sigma$ models. Limiting ourselves to $O(a^2)$ tree improvement we choose 
$c_{n_1n_2n_3n_4}=0$ and the inverse propagator
\begin{equation}
\widetilde{\rho}_{os}(k_0,k_1)=\hat{k}_0^2+\hat{k}_1^2-{1\over 6}
\hat{k}_0^2\hat{k}_1^2=
k_0^2 + k_1^2 - {1\over 12} \left( k_0^2 + k_1^2\right)^2 + O(k^6).
\label{eq32}
\end{equation}
Performing the integrations in the variable $k_0$ we found
\begin{equation}
r_1(L)={1\over 2}-{1\over 3L} \sum_{l=1}^L 
{\hat{k}^2_1\over 2\sqrt{ \hat{k}_1^2 + \case{1}{12}\hat{k}_1^4}}=
0.30755... + O\left( {a^2\over L^2} \right)
\label{eq33}
\end{equation} 
\begin{equation}
r_2(L)={1\over L} \sum_{l=1}^{L-1} 
{1\over 2\sqrt{ \hat{k}_1^2 + \case{1}{12}\hat{k}_1^4}}=
{1\over 2\pi}\ln {L\over a} - 0.0028979... +O\left( {a^4\over L^4} \right)
\label{eq34}
\end{equation} 
Hence only for large $N$ on-shell improvement is effective in reducing
the scaling
violations in the finite-space-volume
mass-gap, even if this is an asymptotic (spectral) property of the model. 
$r_1(L)$ does not behave better than the corresponding 
quantity calculated with the standard action.

A similar pattern we found when considering the action involving
only the quadratic terms of the ``Hasenfratz-Niedermayer''
perfect action. This is not unexpected when we consider the low-momentum
expansion of the corresponding function $\widetilde{\rho}$~\cite{DeHaHaNi}
and realize that it is consistent with $O(a^2)$ on-shell improvement.
In Ref.~\cite{FaHaNiPa} it was verified that the inclusion 
of the ``perfect'' quartic couplings in the action
has the effect of removing all power-law dependence on 
lattice artifacts. Our results for $r_i$ and $s_i$
should now allow to establish the more
general conditions under which, starting from a given form
of the bilinear interaction $\rho_{xy}$
(dictated by the specific RG transformations considered),
one may choose functions $\widetilde{c}$ such that all non-scaling one-loop 
finite lattice effects are removed, at least in the 
finite-space-volume mass gap.

We conclude with a few remarks about four-dimensional lattice QCD.
Similarly to the finite-volume mass gap in 2-d $O(N)$ $\sigma$ models, 
one may define observables
which do not present $O(a^k \ln a)$ terms in their one-loop 
perturbative evaluation even when using the standard Wilson action.
In the following we give an example of such a quantity.
Consider the correlation of two Polyakov lines $P(\vec{r})$ (or equivalently
Wilson loops $T\times R$ with $T\gg R$)
\begin{equation}
C(\vec{r}) = {1\over N} \langle {\rm Tr} P(\vec{r}) P(0)\rangle .
\label{eq35}
\end{equation}
The perturbative expansion of $C(\vec{r})$ has been 
considered in Ref.~\cite{CuMePa} within a study of the Symanzik's 
program in lattice
gauge theories. Using the Wilson action and for an infinite lattice
\begin{equation}
C(\vec{r})=1 + T c_{F} g^2 \int {d^3k\over (2\pi)^3}
{e^{i\vec{k}\cdot\vec{r}} -1 \over \hat{k}_1^2 + \hat{k}_2^2 + \hat{k}_3^2 }
+O(g^4),
\label{eq36}
\end{equation}
where $T$ is the length of the Polyakov line.
As shown in Ref.~\cite{CuMePa}, the one-loop contribution to
\begin{equation}
A(\vec{r})\equiv {1\over T} \ln C(\vec{r}) 
\label{eq37}
\end{equation}
contains $O(g^4 a^{2k}\ln a)$ terms.
This is not unexpected because at tree level 
$O(a^2)$ finite lattice spacing corrections
are present.

Let us now introduce the wall-wall correlation
\begin{equation}
W(z)\equiv {1\over L_x L_y} \sum_{xy} C(\vec{r}).
\label{eq38}
\end{equation}
It is easy to prove that
\begin{equation}
D(z)\equiv {1\over c_{F} T} \ln {W(z+1)\over W(z)}=
g^2 + O(g^4)
\label{eq39}
\end{equation}
similarly to $m(L)L$ in two-dimensional $O(N)$ $\sigma$ models.
Then, unlike $A(\vec{r})$,  the $O(g^4)$ corrections in $D(z)$
will not contain $O(g^4 a^{2k} \ln a)$ terms, and its perturbative
expansion looks
like that of $m(L)L$ with the substitution $z\rightarrow L$.
Notice that the continuum limit of $D(z)$ defines
a running coupling constant  obeying a homogeneous renormalization
group equation
with a corresponding beta-function.

\acknowledgments

It is a pleasure to thank Andrea Pelissetto
for useful and stimulating discussions.



\end{document}